\documentclass[12pt]{article}
\input epsf.sty
\def\ZZZ{{\hbox{ Z\kern-1.6mm Z}}}

\newcommand{\vp}{\varphi}

\newcommand{\tl}{\wt\lambda}

\newcommand{\OO}{{\cal O}}

\newcommand{\LL}{{\cal L}}

\newcommand{\wt}{\widetilde}

\newcommand{\bd}{\bar{\rm D}}

\newcommand{\TT}{{\cal T}}

\newcommand{\SSS}{{\cal S}}

\newcommand{\be}{\begin{equation}}
\newcommand{\ee}{\end{equation}}
\newcommand{\ben}{\begin{eqnarray}\displaystyle}
\newcommand{\een}{\end{eqnarray}}
\newcommand{\refb}[1]{(\ref{#1})}
\newcommand{\p}{\partial}
\newcommand{\sectiono}[1]{\section{#1}\setcounter{equation}{0}}

\def\one{{\hbox{ 1\kern-.8mm l}}}
\def\zero{{\hbox{ 0\kern-1.5mm 0}}}
 
\begin{document}
{}~
{}~
\hfill\vbox{\hbox{hep-th/0312003}
}\break
 
\vskip .6cm
\centerline{\Large \bf Moduli Space of Unstable D-branes}
\medskip
\centerline{\Large \bf   on a Circle of
Critical Radius}

\medskip
 
\vspace*{4.0ex}
 
\centerline{\large \rm
Ashoke Sen}
 
\vspace*{4.0ex}

\centerline{\large \it Harish-Chandra Research
Institute}

\centerline{\large \it  Chhatnag Road, Jhusi,
Allahabad 211019, INDIA}
 
\centerline{E-mail: ashoke.sen@cern.ch,
sen@mri.ernet.in}
 
\vspace*{5.0ex}
 
\centerline{\bf Abstract} \bigskip

We study the moduli space of the boundary conformal field theories
describing an unstable D-brane of type II string theory compactified on a
circle of critical radius. This moduli space has two branches, -- a three
dimensional branch $S^3/Z_2$ and a two dimensional branch described by a
square torus $T^2$. These two branches are joined along a circle. We
compare this with the moduli space of classical solutions of tachyon
effective field theory compactified on a circle of critical radius. This
moduli space has a very similar structure to that of the boundary
conformal field theory with the only difference that the $S^3$ of the
$S^3/Z_2$ component becomes a deformed $S^3$. This provides one more
indication that the tachyon effective field theory captures qualitatively
the dynamics of the tachyon on an unstable D-brane.

\vfill \eject
 
\baselineskip=16.4pt

\tableofcontents

\sectiono{Introduction} \label{s1}

Classical dynamics of unstable D$p$-branes in bosonic and superstring
theories have been investigated using various techniques. A particularly
interesting configuration where many analytical results can be obtained is
when one of the directions of the D-brane is wrapped on a circle of
certain critical radius so that the first non-zero momentum mode of the
tachyon field along the circle becomes a massless mode in the effective
$(p-1, 1)$ dimensional theory. The result of switching on classical 
background value of this massless
mode can be represented as a marginal deformation of the boundary
conformal field theory (BCFT) describing the dynamics of the D-brane
system\cite{others,9811237,9808141,9812031,0108238}. This allows us to 
derive
analytical results about the system at this point. This system is also of
physical interest since by an inverse Wick rotation of the coordinate
along the circle we can transform this configuration to a time dependent
configuration describing the rolling of the tachyon field away from the
maximum of the potential\cite{0203211,0203265,0207105,larsen}.

Since on a circle of critical radius the first momentum mode of the
tachyon becomes a marginal deformation of the conformal field theory, we
have a family of BCFT's labelled by the 
classical vacuum
expectation value of this mode. There are two massless modes, associated
with the two different signs of the momentum along $S^1$. The vacuum
expectation value of these two modes, together with the Wilson line of the
U(1) gauge field on the D-brane along the circle, gives a three parameter
family of BCFT's. The associated three
dimensional moduli space in the case of bosonic string theory is known to
be the $SU(2)$ group manifold $S^3$\cite{9811237}.

In this paper we analyze the moduli space of the conformal field theory
describing an unstable D-brane of superstring theory on a critical circle.
The structure of the moduli space turns out to be more complicated, having
two branches. One of the branches, on which the original D$p$-brane BCFT 
lies, is a $Z_2$ orbifold of $S^3$\cite{0108238}, with
fixed points lying along an equator of $S^3$. 
Physically a BCFT associated with a fixed point of the $Z_2$ 
transformation represents a D-$(p-1)$ $\bd$-$(p-1)$ brane pair, transverse 
to $S^1$, situated at diametrically opposite points on the circle. From 
these fixed points emerges a new branch of the moduli space representing 
D-$(p-1)$-brane $\bd$-$(p-1)$-brane pair placed at {\it arbitrary} points 
on the 
circle. This new branch has the shape of a
square torus, labelled by the positions of the D-$(p-1)$-brane and the 
$\bd$-$(p-1)$-brane on the circle $S^1$. The two branches are joined along 
a circle, with the circle
running along the line of orbifold fixed points in the first branch and
along the diagonal of the square torus in the second branch.

Given that the tachyon field has mass$^2$ of order $-(\alpha')^{-1}$, we
clearly cannot hope to understand the complete tachyon dynamics using an
effective field theory. Nevertheless it has been found that there is a
candidate effective action that describes many features of the dynamics of 
the tachyon
qualitatively\cite{effective,0303139,effective2,effective3}. In particular 
this
action has the property that if we compactify one of the directions on a
circle of critical radius, then there is a family of classical solutions
involving switching on vacuum expectation values of the tachyon and other
fields. We analyze this family of classical solutions and determine the
moduli space of the solutions. It turns out that the moduli space of the
classical solutions has a structure very similar to that of the
BCFT's. In particular it has two branches. 
One of
them can be thought of as a $Z_2$ orbifold of a {\it deformed} $S^3$, with 
a
circle of orbifold fixed points. The other one is exactly a square torus
as in the case of BCFT. These two branches are 
again
joined along a circle, with the circle running along the fixed point of
the $Z_2$ transformation on the deformed $S^3$ and along the diagonal of
the square torus. This is one more indication that the tachyon effective
action of refs.\cite{effective,0303139,effective2,effective3} is capable 
of giving us
qualitative understanding of the full tachyon dynamics in string theory.

The rest of the paper is organized as follows. In section \ref{s2} we
review the construction of the moduli space of BCFT's describing a D-brane 
of bosonic string theory on a critical
circle. In section \ref{s3} we generalize this construction to D-branes in
superstring theory. Section \ref{s4} is devoted to analyzing the moduli
space of classical solutions in the tachyon effective field theory wrapped
on a critical circle. In section \ref{s5} we discuss how the extra moduli
required for the opening up of the other branch $T^2$ arises in the
effective field theory as a result of the zero momentum tachyon mode
becoming massless. Throughout the paper we shall use $\alpha'=1$ unit 
unless mentioned otherwise.

\sectiono{D-string in Bosonic String Theory on a Circle of Unit Radius
} \label{s2}

In this section we review, following \cite{9811237,0108238} (see also
\cite{others}), the moduli space of
the D-brane of bosonic string theory wrapped on a circle of self-dual
radius. 
For definiteness we shall focus our attention on a D-string wrapped on 
$S^1$, and ignore the moduli associated with the motion of the D-string in 
the non-compact transverse directions, but the results are clearly valid 
for all D$p$-branes. If we denote by $x$ the coordinate 
along the compact direction, then the BCFT
associated with the $x$ coordinate admits a three parameter family of
marginal deformations, generated by the
operators $i\partial X$, $\cos X$ and $\sin X$ respectively. The structure
of the moduli space generated by these marginal deformations is best
understood by using the $SU(2)_L\times SU(2)_R$ current algebra that
exists at the self-dual radius. The $SU(2)$ currents are given by:
\ben \label{e1}
J^3_L \sim i\partial X_L, \quad J^1_L \sim \cos (2X_L), \quad J^2_L \sim
\sin
(2 X_L), \nonumber \\
J^3_R \sim i\partial X_R, \quad J^1_R \sim \cos (2 X_R), \quad J^2_R \sim
\sin  
(2 X_R), 
\een
where $X_L$ and $X_R$ denotes the left and the right-moving components of
$X$ so that $X=X_L+X_R$. The boundary condition on $X$ is given by
\be \label{e2}
X_L=X_R\, ,
\ee
which gives
\be \label{e3}
J^a_L=J^a_R, \quad \hbox{for} \quad 1\le a \le 3\, ,
\ee
at the boundary. This shows that the boundary operators $i\p X$, $\cos
X$ and $\sin X$ can be regarded as restrictions of $J^3_L$, $J^1_L$ and
$J^2_L$ (or $J^3_R$, $J^1_R$ and   
$J^2_R$) respectively to the boundary. 

Since $J^a$'s generate $SU(2)$ transformation, the
moduli space of the BCFT associated with the boundary deformations 
generated by $i\p X$, $\cos X$ and $\sin X$
is
simply the $SU(2)$ group manifold $S^3$\cite{9811237,0108238}. We shall
parametrize this manifold by choosing the following representation of the
$SU(2)$ group element:
\be \label{e4}
g = \exp({1\over 2}i \chi \sigma_3) \exp({1\over 2}i \phi \sigma_3) \exp( 
i \theta \sigma_1) \exp(- {1\over 2}i
\phi \sigma_3) \exp({1\over 2}i \chi \sigma_3)\, ,
\ee
where $\sigma_1$, $\sigma_2$, $\sigma_3$ are Pauli matrices.
In this parametrization the metric on $S^3$ is given by:
\be \label{e5}
ds^2 = d\theta^2 + \sin^2\theta d\phi^2 + \cos^2\theta d\chi^2\, .
\ee
Each of $\theta$, $\phi$ and $\chi$ can be taken to be angular variables
with periodicity $2\pi$, provided we make the following 
additional identifications 
\be \label{e6}
(\theta, \phi, \chi) \equiv (2\pi-\theta, \phi+\pi, \chi), \qquad
(\theta, \phi, \chi) \equiv (\pi - \theta, \phi, \chi+\pi)\, .
\ee 
Thus we can take the fundamental region of the moduli space to be:
\be \label{efun1}
0\le\theta\le {\pi/2}, \quad 0\le\phi\le 2\pi, \quad 0\le\chi\le 2\pi\, .
\ee
The $S^3$ nature of the manifold can be made manifest by associating to
$(\theta,\phi,\chi)$ the point $(\cos\theta\cos\chi, \cos\theta\sin\chi,
\sin\theta\cos\phi, \sin\theta\sin\phi)$ in $R^4$. These points clearly
span a unit three sphere in $R^4$.

It is instructive to try to identify the parameters $(\theta,\phi,\chi)$ 
with the explicit
boundary deformations generated by $i\partial X$, $\cos X$ and $\sin X$.
For this we begin with the observation that a generic boundary 
perturbation of the form $-i\vp^a \ointop J^a_L$ for some parameters 
$\vp^a$, with $\ointop J^a_L$ denoting $(2\pi i)^{-1}$ times an integral 
of $J^a_L(z)$ over $z$ along 
the boundary of the 
world-sheet, corresponds to an insertion of $\exp(i\vp^a \ointop J^a_L)$ 
in the euclidean world-sheet path integral.
(The choice of $SU(2)_L$ is purely a
matter of convention; we could have also chosen the $SU(2)_R$ group since
the boundary condition $J_L^a=J_R^a$ relates the two sets of generators.)
Thus any BCFT associated with deformation 
by 
the operators $i\partial X$, $\cos X$ and $\sin X$ can be associated with 
an $SU(2)_L$ group element $\exp(i\vp^a \ointop J^a_L)$. We can now make 
a natural association between 
the 
$SU(2)$ group element $g$ given in \refb{e4} and an $SU(2)_L$ group 
element 
by associating 
${1\over 2}\sigma_a$ with the generators
$\ointop J^a_L$.
Thus the 
element $g$ given in \refb{e4} is associated with the following insertion 
in the path integral:
\be \label{e6.1}
\exp(i \chi \ointop J^3_L) \exp(i \phi \ointop J^3_L) \exp(
2i \theta \ointop J^1_L) \exp(- i
\phi  \ointop J^3_L) \exp(i \chi\ointop J^3_L)\, .
\ee
This can be interpreted as the insertion of a series of exponentiated 
contour integrals, with the left-most element containing the outermost 
contour 
closest to the world-sheet boundary and the rightmost element containing 
the 
innermost contour. Now the $\ointop J^3_L$ in the left-most contour can be 
converted into a $-\ointop J^3_R$ using the boundary condition \refb{e3}, 
the change in sign being due to the change in the direction of 
integration. Since 
$J^a_R$ commute with $J^b_L$, this element can now be taken all the way to 
the right and combined with the last term to give $\exp(i \chi\ointop 
(J^3_L-J^3_R))$. In this new configuration $\exp(i \phi \ointop J^3_L 
)$ becomes the left-most term, and hence again the $\ointop J^3_L$ in this 
term 
can be converted into a $-\ointop J^3_R$ using the boundary condition 
\refb{e3}. 
Again using the fact that $J^a_R$ commutes with $J^b_L$, we can bring this 
term to the right of $\exp(
2i \theta \ointop J^1_L)$ and combine it with the $\exp(-i
\phi  \ointop J^3_L)$ term. Thus the net result is:
\be \label{e6.2}
\exp(
2i \theta \ointop J^1_L) \exp(- i
\phi  \ointop (J^3_L+J^3_R)) \exp(i \chi\ointop (J^3_L - J^3_R))\, .
\ee

This has the following interpretation. The insertion of $\exp(
2i \theta \ointop J^1_L)$ corresponds to deforming the BCFT by a term 
proportional to $\cos X$. The insertion of $\exp(- 
i
\phi  \ointop (J^3_L+J^3_R))$ corresponds to translating 
the world-sheet field $X$ by an amount proportional to $\phi$,
since $\ointop (J^3_L+J^3_R)$ is the generator of ordinary translation. 
This can also be seen explicitly by rewriting the piece $\exp(i \phi 
\ointop J^3_L) \exp(
2i \theta \ointop J^1_L) \exp(- i
\phi  \ointop J^3_L)$ in the original expression \refb{e6.1}
as $\exp(
2i \theta \ointop (J^1_L\cos\phi - J^2_L \sin\phi))$. This, using 
\refb{e1}, 
corresponds to a boundary deformation proportional to $\cos(X+\phi)$.
Thus the effect of $\phi$ is to
essentially convert the $\cos X$ perturbation to 
$\cos(X+\phi)$. 
Finally, the effect of $\exp(i \chi\ointop (J^3_L 
- J^3_R))$ is to induce a translation proportional to $\chi$ along the 
T-dual circle, since $\ointop (J^3_L 
-
J^3_R)$ is the generator of translation along the T-dual circle. 
Since the operators $\ointop(J^3_L\pm J^3_R)$ multiplying $\phi$ and 
$\chi$ are 
normalized in the same way, and since $\phi$ is precisely the amount of 
translation on the original circle, $\chi$ must measure precisely the 
amount of translation along $S^1_D$. This is also consistent with the fact 
that 
$\chi$ appearing in \refb{e4}, as well as the 
T-dual 
circle $S^1_D$, has periodicity $2\pi$.
In the 
original description switching on $\chi$ corresponds to inducing a Wilson 
line $\chi$ 
along 
$S^1$, normalized so as to have periodicity $2\pi$.

For later use, we would like to find the constant of proportionality 
between the parameter 
$\theta$ and the coefficient $\tl$ of the $\cos X$ boundary perturbation 
that appears {\it e.g.} in \cite{0203211,0203265}. For this
we
note that at $\theta=0$ the orbits generated by $\phi$ translation shrink
to points. Since $\phi$ represents translation along $S^1$, this implies
that the $\theta=0$ configuration is invariant under translation along
$S^1$. This is of course consistent with the identification of the
$\theta=0$ point with a D-string wrapped on $S^1$ which is translationally
invariant. We also see from \refb{e5} that at $\theta=\pi/2$
the orbits generated by $\chi$ translation shrink to points. Since
$\chi$-translation corresponds to translation along the dual circle
$S^1_D$, this shows that the $\theta=\pi/2$ point must represent a
configuration of the D-brane which is invariant under translation along
the dual circle. Thus it is natural to identify the $\theta=\pi/2$ point
as the point where the original D-string is converted to a D0-brane, 
{\it
i.e.} the $\tl={1\over 2}$ point in the notation of \cite{0203211}. 
Since T-duality converts this to a D-string wrapped on a dual circle,
this configuration is natuarally invariant under translation along the 
dual circle.
Thus the
parameter $\theta$ is related to the parameter $\tl$ of \cite{0203211} 
through
the relation:
\be \label{e7} 
\theta = \pi \tl\, .
\ee

To summarize, in order to associate a BCFT
corresponding to a given point $(\theta,\phi,\chi)$ on the $SU(2)$ group
manifold, we proceed as follows. First we switch on a boundary deformation
proportional to $\theta\cos(X+\phi)$ with the overall normalization
adjusted so that $\theta=\pi/2$ corresponds to the `Dirichlet point'. We
then go to the T-dual description and translate the BCFT
by an amount $\chi$ on the T-dual circle. Finally we go back
to the original description by making a further T-duality transformation.

Let us now check that the metric \refb{e5} on
the $SU(2)$ group manifold agrees with the metric on the moduli 
space of classical solutions in open string field theory for small
$\theta$. For this we consider the quadratic part of the open string field
theory
action describing small fluctuations of the tachyon field $T$ and the 
gauge field $A_x$ on the D-string wrapped on $S^1$:
\be \label{e8}
S^{(2)} = -\TT_1\, \int \, dt \, \int_0^{2\pi} \, dx \, \left[1 - {1\over 
2} T^2 
- 
{1\over 2} \dot T^2 + {1\over 2} (T')^2 - {1\over 2} (\dot A_x)^2\right]\, 
,
\ee
where $\TT_1$ denotes the tension of the D-string. The constant term in 
the action denotes the contribution due the tension of the D-string.
The gauge field component $A_x$ is an angular variable, since it measures 
the Wilson line along $S^1$. To determine the 
periodicity of $A_x$, we begin with the observation that 
for $T=0$, $A_x=0$, the total energy of the brane is $2\pi\TT_1$. In the 
T-dual description this can be interpreted as the mass $\TT_0$ of the 
D0-brane on $S^1_D$. We now note that for $x$ independent $A_x$, the term 
involving $\dot A_x$ in \refb{e8} has the form
\be \label{e9}
{1\over 2}\, \TT_0\, \int \, dt \, (\dot A_x)^2\, .
\ee
This can be interpreted as the action for the D0-brane moving on
$S^1_D$ if we interprete $A_x$ as the position of the D0-brane 
on $S^1_D$.
Since $S^1_D$ has unit radius, we see that 
$A_x$ is an angular coordinate with periodicity $2\pi$.

Consider now a classical soliton solution of this field theory given by
\be \label{e10}
T = \lambda \cos x \, .
\ee
The massless fluctuations around this solution are associated with a 
general field configuration of the form:
\be \label{e11}
T(x,t) = \lambda(t) \cos (x + \alpha(t))\, , \qquad A_x(x,t) = 
\beta(t)\, ,
\ee
where $\lambda$, $\alpha$ and $\beta$ are collective coordinates. From 
\refb{e11} it is clear that $\alpha$ is an angular variable 
with 
periodicity $2\pi$. Since $A_x$ is an angular variable
with
periodicity $2\pi$, it also follows from \refb{e11} 
that $\beta$ is an angular 
variable
with
periodicity $2\pi$.
Substituting \refb{e11} into the action \refb{e8} we get the 
action for the collective coordinates:
\be \label{e12}
S^{(2)} = - \TT_0 \, \int \, dt \, [1 - {1\over 4} \dot\lambda^2 - {1\over 
4} 
\lambda^2 \dot\alpha^2 - {1\over 2} \dot\beta^2 ]\, .
\ee
We want to identify the collective coordinates $\lambda$, $\alpha$ and 
$\beta$ with the coordinates $\theta$, $\phi$ and $\chi$ labelling the 
$SU(2)$ moduli space near $\theta=0$. 
Since $\alpha$ generates translation along $S^1$, we have
$\alpha=\phi$. $\beta$, being the 
Wilson line along $S^1$, generates 
translation along the dual circle $S^1_D$. Thus $\beta$ must be
proportional to $\chi$. Since both $\chi$ and $\beta$ have periodicity 
$2\pi$, we have
$\beta=\chi$. Finally
we need to find the relation between the 
parameter $\lambda$ labelling the solution in the open string field theory 
and the parameter $\theta$ (or equivalently the parameter 
$\wt\lambda=\theta/\pi$) 
labelling the solution in the BCFT. For 
this we 
compare the $xx$ component of the stress tensor in the two 
descriptions.\footnote{Since $T_{xx}$ acts as the source for the zero 
momentum closed string sector scalar field $g_{xx}$, this is a physical 
quantity. Thus it 
is appropriate to match the parameters in the two descriptions by 
comparing the $T_{xx}$ for the two systems.}
In the open string
field theory we have for the solution \refb{e10},
\be \label{e13}
T_{xx} = - \TT_1  \left(1 - {1\over 2} (T')^2 -{1\over 2} T^2\right ) = 
-\TT_1( 1 - {1\over 2} \lambda^2) \, .
\ee
On the other hand for the BCFT
the $T_{xx}$ associated with the  
boundary deformation
$\wt\lambda\cos(X)$ is obtained by Wick rotation of the 
answer for $T_{00}$ associated with the boundary 
deformation $\tl\cosh(X^0)$. This gives\cite{0203265}: 
\be \label{e14}
T_{xx} = - \TT_1 \cos^2 (\pi \tl) = -\TT_1 \cos^2\theta \simeq -\TT_1 
(1-\theta^2)\, 
,
\ee
for small $\tl$ {\it i.e.} small $\theta$. Comparing \refb{e13} and 
\refb{e14} we see that for small $\lambda$ we must have the 
identification $\lambda\simeq\sqrt 2\theta$. Thus all in all, we have:
\be \label{e15}
\alpha=\phi, \quad \beta=\chi, \quad \lambda \simeq \sqrt 2 \theta\, ,
\ee
Substituting this into \refb{e12} we get
\be \label{e16}
S^{(2)} \simeq - \TT_0 \, \int \, dt \, [1 - {1\over 2} \dot\theta^2 - 
{1\over 
2}
\theta^2 \dot\phi^2 - {1\over 2} \dot\chi^2 ]\, .
\ee
This corresponds to a moduli space metric:
\be \label{e17}
ds^2 \simeq (d\theta^2 + \theta^2 d\phi^2 + d\chi^2) \, ,
\ee
near $\theta=0$.
This agrees with the exact metric \refb{e5} near $\theta=0$.

\sectiono{Unstable D-string of Superstring Theory on a Circle of
Radius $\sqrt{2}$} \label{s3}

In this section we shall repeat the analysis of the previous section for 
an unstable D-brane in superstring theory, wrapped on a circle 
$S^1$
of radius $\sqrt{2}$, and show that it has the structure of 
$S^3/Z_2$\cite{0108238} and 
$T^2$ joined together along a circle. 
Again for definiteness we shall focus on the case of unstable D-string of 
type IIA string theory and ignore the moduli associated with the 
transverse  motion of the D-string, but the results are valid for any 
unstable D-brane in type IIA or IIB string theory. 
Let $x$ denote the coordinate along the circle, $X$ be the associated 
world-sheet scalar field, and $\psi$ be the world-sheet superpartner of 
$X$. At the critical radius $\sqrt 2$ the BCFT
associated with the fields $(X,\psi)$ admits three marginal boundary 
deformations\cite{9808141,9812031}.\footnote{In the notation used later in 
this section, \cite{9812031} describes the marginal deformation of the 
theory from the $\theta=0$ point to the $\theta = \pi/2$ point, whereas 
\cite{9808141} describes the marginal deformation of the 
theory from the $\theta=\pi/2$ point to the $\theta = 0$ point.} Up to 
Chan-Paton factors, these operators are
\be \label{e2.1}
i\partial X, \quad \psi \sin (X/\sqrt 2), \quad -\psi \cos(X/\sqrt 2)\, .
\ee
Here we have used a zero picture representation\cite{FMS}.
These represent respectively the effect of switching on a Wilson line 
along $S^1$, switching on a tachyon background proportional to 
$\cos(x/\sqrt 2)$ and  a tachyon background proportional to
$\sin(x/\sqrt 2)$.
In order to analyze the BCFT's
generated by these marginal deformations, we use a fermionic 
representation for the bosonic field $X$. We have, up to cocycle and 
numerical  
factors\cite{9808141,9812031}:
\be \label{e2.2}
e^{\pm i \sqrt 2 X_L} \sim (\xi_L \pm i\eta_L), \quad e^{\pm i\sqrt 2 X_R} 
\sim (\xi_R \pm 
i\eta_R), \quad i\partial X_L \sim \eta_L\xi_L, \quad i\partial X_R \sim 
\eta_R\xi_R\, ,
\ee
where $(\xi_L, \eta_L)$ denote a pair of left-handed Majorana-Weyl fermion 
and 
$(\xi_R, \eta_R)$ denote a pair of right-handed Majorana-Weyl fermion. 
The 
Neumann boundary condition $X_L=X_R$, $\psi_L=\psi_R$ translates to:
\be \label{e2.3}
\xi_L=\xi_R\equiv\xi, \quad \eta_L=\eta_R\equiv \eta, \quad 
\psi_L=\psi_R\equiv\psi\, .
\ee
Thus the conformal field theory described by $(X,\psi)$ can be equally 
well described as a theory of three free fermion fields $\psi$, $\xi$ and 
$\eta$. This theory has an underlying level two $SU(2)_L\times SU(2)_R$ 
current 
algebra, generated by:
\be \label{e2.3a}
J^1_L \sim  \psi_L\eta_L, \quad J^2_L \sim  \xi_L\psi_L, \quad J^3_L 
\sim 
\eta_L\xi_L, \quad J^1_R \sim  \psi_R \eta_R, \quad J^2_R \sim 
\xi_R\psi_R, 
\quad J^3_R \sim
\eta_R\xi_R\, .
\ee
The boundary condition \refb{e2.3} gives
\be \label{e2.3b}
J^a_L=J^a_R 
\ee
on the boundary. The existence of this current algebra does not however 
mean that the underlying theory is invariant under this symmetry. For 
example, in the 
closed string sector the GSO projection rules do not commute with 
the $SU(2)_L$ and $SU(2)_R$ symmetry. In the open string sector the 
spectrum is invariant 
under the diagonal $SU(2)$ group, but the correlation functions are not 
$SU(2)$ invariant since the assignment of Chan Paton and cocycle factors 
break this $SU(2)$ symmetry\cite{9808141,9812031}. Nevertheless we
shall see that the $SU(2)$ 
symmetry can be exploited to determine the structure of the moduli space.

Using eqs.\refb{e2.2} - \refb{e2.3b}, and that
$X=X_L+X_R$, the three boundary operators listed in eq.\refb{e2.1} may be 
expressed as:
\be \label{e2.4}
i\p X\sim \eta_L\xi_L\sim J^3_L, \quad \psi\sin(X/\sqrt 2) 
\sim \psi_L\eta_L \sim J^1_L, \quad 
-\psi\cos(X/\sqrt 2) \sim \xi_L\psi_L 
\sim J^2_L\, .
\ee
Had $SU(2)$ been an exact symmetry of the problem, this would again imply 
that the moduli space of the BCFT generated 
by 
these marginal deformations is locally $SU(2)$ group manifold $S^3$. As we 
have pointed out, $SU(2)$ is not a symmetry of the full string theory in 
this case.
However for the 
marginal operators listed in \refb{e2.4}, the corresponding Chan-Paton 
times cocycle factors commute (being identity for the first operator and 
the Pauli matrix
$\Sigma_1$ for the second and the third operator\cite{9812031}), and hence
plays no role 
in the computation of the correlation function of these operators. Since 
the computation of the moduli space metric only involves computing 
correlation functions of these operators (together with inverse picture 
changing 
operators\cite{inverse} which are manifestly $SU(2)$ invariant), we see
that as far as 
computation of the moduli space metric is concerned, we can treat the 
$SU(2)$ symmetry to be unbroken. The marginal deformations generated by 
the boundary values of $J^a_L$ ($1\le a\le 3$) then generate the $SU(2)$ 
group manifold $S^3$, at least locally. As we shall see shortly, global 
structure of the moduli space differs from that of the $SU(2)$ group 
manifold in a subtle way\cite{0108238}.

As in the previous section, we shall label the $SU(2)$ group element as
\be \label{e2.5}
g = \exp({1\over 2}i \chi \sigma_3) \exp({1\over 2}i \phi \sigma_3) \exp( 
i \theta \sigma_1) \exp(- {1\over 2}i
\phi \sigma_3) \exp({1\over 2}i \chi \sigma_3)\, ,
\ee
so that the metric on $S^3$ takes the form:
\be \label{e2.6}
ds^2 = 2(d\theta^2 + \sin^2\theta d\phi^2 + \cos^2\theta d\chi^2)\, .
\ee
The overall normalization factor of 2 reflects that we have a level two 
current algebra.
We can associate a specific BCFT to the 
group 
element $g$ given in \refb{e2.5} by following the same procedure as in the 
case of bosonic string theory. As before, 
the effect of switching on $\theta$ is to switch on a marginal boundary 
deformation 
proportional to $J^1_L\sim \psi\sin(X/\sqrt 2)$, which in turn represents 
a tachyon background proportional to $\cos(X/\sqrt 2)$.
Up 
to an overall normalization $\phi$ and $\chi$ represent respectively the 
effect of translating the BCFT along the 
original circle $S^1$ of radius $\sqrt 2$ 
and along the dual circle $S^1_D$ of radius $1/\sqrt 2$. 
In particular, the effect of switching on $\phi$ is to rotate $\sigma_1$ 
to $\sigma_1 \cos\phi - \sigma_2\sin\phi$, and hence $J^1$ to $J^1\cos\phi 
- J^2\sin \phi\sim \psi\sin({X\over \sqrt 2} + \phi)$. This corresponds to 
translation of $X$ by $\sqrt 2 \phi$. 
{}From this we can conclude
that $\chi$ will induce a translation along $S^1_D$ by an amount $\sqrt 2 
\chi$ 
since the operators multiplying $\chi$ and $\phi$, -- 
$\ointop(J^3_L\pm J^3_R)$, -- are normalized in the same way.
Finally, the relative normalization between $\theta$ and the 
parameter $\tl$ used in \cite{0203265} is determined as follows.
The collapse of the $\chi$ orbit at $\theta=\pi/2$ implies that the 
corresponding configuration must be invariant under translation along the 
dual circle $S^1_D$. Thus this must correspond to the point $\tl=1/2$ in 
the notation of ref.\cite{0203265} where the tachyon background converts 
the D1-brane into a D0-$\bd$0-brane pair situated at the diametrically 
opposite points of the circle. Under T-duality this maps to a D1- 
$\bd$1-brane pair wrapped on $S^1_D$ with half unit of Wilson line on one 
of them, and hence is invariant under translation along $S^1_D$. This
gives 
\be \label{ethla}
\theta=\pi\tl\, .
\ee

Since $S^1_D$ has radius $1/\sqrt 2$, and $\chi$ induces a translation 
along $S^1_D$ by $\sqrt 2 \chi$, we see that $\chi$ has periodicity $\pi$. 
On the other hand, $\chi$ labelling the $SU(2)$ group manifold $S^3$ has 
periodicity $2\pi$. 
This shows that the moduli space of BCFT's,
instead of being an $SU(2)$ group manifold $S^3$, is actually a $Z_2$ 
quotient of $S^3$ by the transformation
$\chi\to\chi+\pi$\cite{0108238}. 
We can see the physical origin of the identification under 
$\chi\to\chi+\pi$
by
working near the point
$\theta=\pi/2$, {\it i.e.} $\tl=1/2$. 
This configuration represents a D0-$\bd$0-brane pair at
diametrically opposite points on the circle $S^1$\cite{9808141,9812031}.
In the T-dual version we have a D1-$\bd$1-brane pair on the dual circle
$S^1_D$ of radius $1/\sqrt 2$, with half a unit of Wilson line along one
of the D-branes. As a
result the tachyonic modes in the open string sector with one end on the
D-string and the other end on the $\bd$-string are forced to carry
half-integral units of momentum $(n+{1\over 2})\sqrt 2$. The lowest modes,
carrying momenta $\pm{1\over \sqrt 2}$, are massless. 
Deforming $\theta$ away from $\pi/2$ corresponds to switching on vacuum
expectation values of some of these modes of the form $({1\over
2}-\tl)\cos({x_D\over \sqrt2}+\chi)\sim ({\pi\over 2} -
\theta) \cos({x_D\over \sqrt 2}+\chi)$, $x_D$ being the coordinate 
labelling
the dual circle. Since $S^1_D$ has radius $1/\sqrt 2$, any given
configuration must go into a physically equivalent configuration under
$x_D\to x_D + 2\pi/\sqrt 2$.
However under this translation $({\pi\over 2} -\theta)\cos({x_D\over \sqrt
2}+\chi)$ 
changes sign rather
than remaining invariant. 
This shows that
we must identify the pair of
solutions $\pm ({\pi\over 
2}-\theta)\cos({x_D\over \sqrt2}+\chi)$, {\it i.e.} identify the points
$(\theta,\phi,\chi)$ and $(\pi-\theta, \phi, \chi)$, or equivalently, the 
points $(\theta,\phi,\chi)$ and $(\theta,\phi,\chi+\pi)$. 
The resulting manifold can be parametrized by
$(\theta,\phi,\chi)$ with the identification:
\be \label{e3.iden}
(\theta,\phi,\chi) \equiv (\theta+2\pi,\phi,\chi) \equiv 
(\theta,\phi+2\pi,\chi) \equiv
(\theta,\phi,\chi+\pi) \equiv (2\pi-\theta,\phi+\pi,\chi) \equiv
(\pi -\theta,\phi,\chi)\, .
\ee
This, in turn, allows us to choose the 
fundamental 
region labelling the moduli space to be:
\be \label{eextra}
0\le\theta\le {\pi\over 2}, \quad 0\le\phi\le 2\pi, \quad 0\le \chi\le 
\pi\, .
\ee

To summarize the results obtained so far, we have the following 
identification between a point 
$(\theta,\phi,\chi)$ on $S^3/Z_2$ and a deformation of the
BCFT describing
an unstable D-brane. We first switch on a boundary deformation 
proportional to 
$\theta \psi \sin({X\over \sqrt 2} + \phi)$, normalized so that 
$\theta={\pi\over 2}$ corresponds to the `Dirichlet point'. We then go to 
the T-dual description of this conformal field theory and translate the 
BCFT by an amount $\sqrt 2\chi$ along $S^1_D$. Finally we go back 
to the original description by making a further T-duality transformation.

We shall now repeat the analysis described at the end of section 
\ref{s2} to check that the moduli space metric \refb{e2.6}
is compatible with the metric on the moduli space of classical solutions 
in open superstring field theory for small $\theta$.
In this case the quadratic part of the string field theory
action describing the dynamics of the tachyon field $T$ and the gauge 
field $A_x$ on the D-string wrapped on $S^1$ is given by:
\be \label{e2.7}
S^{(2)} = -\TT_1\, \int \, dt \, \int_0^{2\pi\sqrt 2} \, dx \, \left[1 - 
{1\over 4} 
T^2
-
{1\over 2} \dot T^2 + {1\over 2} (T')^2 - {1\over 2} (\dot A_x)^2\right]\,
,
\ee
where $\TT_1$ denotes the tension of the non-BPS D-string.
The factor of $1/4$ in front of $T^2$ reflects that the tachyon on a 
non-BPS D-brane in superstring theory has mass$^2=-1/2$.
At $T=0$,
$A_x=0$ the total mass of the system is given by $2\pi\sqrt 2 \TT_1$.
Under a T-duality the bulk theory gets mapped to type IIB string
theory compactified on a dual circle of radius $1/\sqrt 2$, and the
D-string
gets mapped
to a D0-brane of type IIB theory with mass $\TT_0=2\pi\sqrt 2 \TT_1$. Thus
the term involving $A_x$ can be written as
\be \label{e2.9}
{1\over 2}\, \TT_0\, \int \, dt \, (\dot A_x)^2\, .
\ee
Identifying this with the kinetic term for the D0-brane associated with
its motion along $S^1_D$, we see that
$A_x$ denotes the position of the D0-brane on $S^1_D$. Thus $A_x$ is an
angular variable with periodicity $2\pi/\sqrt 2$.

We now consider a classical soliton solution of this field theory given by
\be \label{e2.10}
T = \lambda \cos (x/\sqrt 2) \, .
\ee
The massless fluctuations around this solution are associated with a 
general field configuration of the form:
\be \label{e2.11}
T(x,t) = \lambda(t) \cos (x/\sqrt 2 + \alpha(t))\, , \qquad A_x(t) = 
\beta(t) / \sqrt 2\, ,
\ee
where $\lambda$, $\alpha$ and $\beta$ are collective coordinates. 
With this normalization,
$\alpha$ and $\beta$ are both angular variables 
with 
periodicity $2\pi$.
Substituting \refb{e2.11} into the action \refb{e2.7} we get the 
action for the collective coordinates:
\be \label{e2.12}
S^{(2)} = - \TT_0 \, \int \, dt \, [1 - {1\over 4} \dot\lambda^2 - {1\over 
4} 
\lambda^2 \dot\alpha^2 - {1\over 4} \dot\beta^2 ]\, .
\ee
We need to relate the collective coordinates $\lambda$, 
$\alpha$ and 
$\beta$ with the coordinates $\theta$, $\phi$ and $\chi$ labelling the 
$SU(2)$ moduli space near $\theta=0$.
Since both $\alpha$ and $\phi$ generate translation along $S^1$ and both 
have periodicity $2\pi$, we must have $\alpha=\phi$. On the other hand, 
both $\beta$ and $\chi$ generate translation along the dual circle 
$S^1_D$, but $\beta$ has periodicity 
$2\pi$ whereas $\chi$ has periodicity $\pi$. 
Thus we have $\beta=2\chi$. Finally, 
to find the relation between the parameter $\lambda$
and the parameter $\theta$ (or equivalently the parameter 
$\wt\lambda=\theta/\pi$) 
labelling the BCFT, we
compare the $xx$ component of the stress tensor in the two descriptions.
In the string
field theory the $xx$ component of the stress tensor associated with the 
solution $T=\lambda 
\cos(x/\sqrt 2)$ is
\be \label{e2.13}
T_{xx} = -\TT_1  (1 - {1\over 4} T^2 - {1\over 2} (T')^2) = -\TT_1\, (1 
- {1\over 4} \lambda^2) \, .
\ee
On the other hand $T_{xx}$ associated with the BCFT
obtained by deforming the world-sheet action by a boundary term 
proportional to $\wt\lambda\psi\sinh(X/\sqrt 2)$ is obtained by 
Wick 
rotating the answer for $T_{00}$ associated with the boundary deformation 
$\wt\lambda\psi^0\sinh(X^0/\sqrt 2)$. This can be read out from
ref.\cite{0203265} and gives:
\be \label{e2.14}
T_{xx} = -\TT_1 \cos^2 (\pi \tl) = -\TT_1 \cos^2 \theta \simeq 
-\TT_1
(1-\theta^2)\, 
,
\ee
for small $\theta$. Comparison of \refb{e2.13} and 
\refb{e2.14} gives, for small $\theta$, $\lambda \simeq 2 \theta$. Thus 
the 
complete relation between $(\lambda,\alpha,\beta)$ and 
$(\theta,\phi,\chi)$ for small $\theta$ is given by:
\be \label{e2.15}
\alpha=\phi, \quad \beta=2\chi, \quad \lambda \simeq 2 \theta\, .
\ee 
Substituting this into \refb{e2.12} we get
\be \label{e2.16}
S^{(2)} \simeq - \TT_0 \, \int \, dt \, [1 -  \dot\theta^2 -
\theta^2 \dot\phi^2 - \dot\chi^2 ]\, .
\ee
This corresponds to the metric:
\be \label{emmet}
ds^2 \simeq 2(d\theta^2 + \theta^2 d\phi^2 + d\chi^2)\, ,
\ee
and
agrees with \refb{e2.6} near $\theta=0$.

The above analysis confirms our earlier assertion that the moduli space of 
BCFT's associated with marginal boundary 
deformation of an unstable D-brane wrapped on a critical circle is 
$S^3/Z_2$. However this is only one branch of the moduli space. At
the point $\theta=\pi/2$ there are other branches of the moduli
space of the BCFT. To see this we go to the
T-dual picture where the $\theta=\pi/2$ point represents a
D1-$\bd$1-brane pair, each wrapped on a dual circle $S^1_D$, with the
$\bd$1-brane carrying half a unit of Wilson line. This configuration
supports\cite{9808141} 
six massless modes, associated with the (zero picture) vertex operators:
\ben \label{e2.18}
&& i\p X_D, \quad i\p X_D\otimes \sigma_3, \quad \psi_D \sin {X_D
\over \sqrt 2}\otimes
\sigma_1, \quad  -\psi_D \cos {X_D
\over \sqrt 2} \otimes 
\sigma_1, \nonumber \\
&& \quad \psi_D \sin {X_D
\over \sqrt 2} \otimes 
\sigma_2, \quad  -\psi_D \cos {X_D
\over \sqrt 2} \otimes
\sigma_2\, ,
\een
where the Pauli matrices $\sigma_i$ label the Chan-Paton factors, and
$\psi_D$ and $X_D$ denote the T-dual of $\psi$ and $X$
respectively. The coefficient of $i\p X_D\otimes \sigma_3$ measures the
relative Wilson line between the D1-brane and the $\bd$1-brane. Switching
on this Wilson line makes all the other modes except $i\p X_D$ massive or
tachyonic. Thus there is a two dimensional branch of the moduli space,
labelled by the coefficients of the operators $i\p X_D$ and $i\p
X_D\otimes \sigma_3$, emanating from $\theta=\pi/2$. In the original
description a point on this two 
dimensional branch represents a D0-$\bd$0 brane pair placed at two 
arbitrary
points 
on the 
circle $S^1$. Thus the moduli space has the structure of a square torus 
$T^2=S^1\times S^1$, with flat metric. This branch has a one
dimensional intersection with the earlier branch $S^3/Z_2$.
On the square torus, this one dimensional line runs along the diagonal,
representing D0-$\bd$0-brane pairs at diametrically opposite points on the
circle. On $S^3/Z_2$, this line is the equator labelled by $\phi$,
situated at $\theta=\pi/2$.

If we set the coefficient of the operator $i\p X_D\otimes \sigma_3$ to 
zero, 
we can switch on vev of the other massless fields. 
However we cannot simultaneously switch on all of the five other massless 
fields, since 
in order to have a
mutually marginal set of operators we need to have operators with
commuting Chan-Paton factors. This leaves us with a one parameter ($\tau$) 
family
of three dimensional branches,  generated by the operators:
\be \label{e2.19}
i\p X_D, \quad \psi_D \sin {X_D\over \sqrt 2} \otimes
(\cos\tau \sigma_1 + \sin\tau \sigma_2), \quad 
-\psi_D \cos {X_D\over \sqrt 2} \otimes(\cos\tau \sigma_1 + \sin\tau
\sigma_2)\, .
\ee
The operators for different values of $\tau$ however are related to each
other by a $U(1)$ gauge symmetry on the world volume of the
D1-$\bd$1 system, generated by $\sigma_3$. Since in constructing the
moduli space of solutions we must identify solutions which are related by
open-string gauge transformations, we see that there really is just 
one three
dimensional branch. We can take this to be the branch corresponding to
$\tau=0$, {\it i.e.} the branch already discussed earlier. This 
argument also shows that the solutions associated with $\tau=0$ and 
$\tau=\pi$ must
be
identified. This changes the sign of the operators $\psi_D \cos X_D
\otimes 
\sigma_1$ and $\psi_D \sin X_D \otimes
\sigma_1$, and precisely corresponds to the $\chi\to\chi+\pi$
transformation discussed earlier. Thus we see that the origin of the $Z_2$
quotient can be traced back to a residual $Z_2$ subgroup of
the $U(1)$ gauge symmetry on the D1-$\bd$1 brane system.

To summarize, the moduli space of the BCFT's
associated with a non-BPS D-string wrapped on a circle of radius $\sqrt 2$ 
has two branches, -- a three dimensional branch $S^3/Z_2$, and a two 
dimensional branch with the geometry of a square torus $T^2$. These two 
branches are joined along a circle. On the torus this circle 
runs along the diagonal of the square torus, whereas on $S^3/Z_2$ the 
circle runs along the line of orbifold fixed points situated at 
$\theta=\pi/2$.

\sectiono{Moduli Space of Solutions in the Tachyon Effective Field Theory}
\label{s4}

Many of the qualitative (and some quantitative) features of the tachyon
dynamics on an unstable D$p$-brane are described by the tachyon effective
action\cite{effective,0303139,effective2,effective3}:
\ben \label{ez1}
S &=& \int d^{p+1} x \, \LL\, , \nonumber \\
\LL &=& - V(T) \, \sqrt{-\det A}
\, ,
\een
where
\be \label{ey2}
A_{\mu\nu} = \eta_{\mu\nu} + \p_\mu T \p_\nu T + \p_\mu Y^I \p_\nu Y^I + 
F_{\mu\nu}\, ,
\ee
\be \label{ey2a}
F_{\mu\nu} = \p_\mu A_\nu - \p_\nu A_\mu\, .
\ee
$A_\mu$ and $Y^I$ for $0\le \mu, \nu\le p$, $(p+1)\le I\le 9$ are the
gauge and the transverse scalar fields on the world-volume of the non-BPS
brane, and $T$ is the tachyon field.
$V(T)$ is the tachyon potential:
\be \label{ess1}
V(T) = {\TT_p \over \cosh(T/\sqrt 2)}\, .
\ee
We shall focus on the case $p=1$ and take the coordinate $x^1=x$ to be
along a circle $S^1$ of radius $\sqrt 2$. 
We shall also ignore the dynamics of the transverse coordinates $Y^I$
since they simply represent transverse motion of the brane and can be
incorporated at  a later stage if needed. In this case the action
\refb{ez1} in the $A_0=0$ gauge reduces to:
\be \label{ess1a}
S = -\TT_1 \, \int \, dt \, \int_0^{2\pi\sqrt 2} \, dx \, {1\over
\cosh(T/\sqrt 2)} \sqrt{1 - \dot T^2 + (T')^2 - (\dot A_x)^2}\, .
\ee 
Since 
for small $T$ and $A_x$
the action \refb{ez1} reduces to the quadratic action
\refb{e2.7}, the arguments of section \refb{s3} can be used to
conclude that $A_x$ has periodicity $2\pi/\sqrt 2$.

The
theory
described by the action \refb{ess1a} has a three parameter family of
classical solutions\cite{0303139}:
\be \label{ess2}
T = \sqrt 2 \sinh^{-1} \left({\lambda\over \sqrt 2} \cos\left({x\over 
\sqrt 2} +
\alpha\right)\right), \quad A_x = {1\over \sqrt 2} \beta\, .
\ee
Here $\lambda$, $\alpha$ and $\beta$ are the
parameters labelling the solution with $\alpha$ and $\beta$ being angular 
variables with periodicity $2\pi$. 
Our goal is to find the metric in the three parameter moduli 
space
labelled by $\lambda$, $\alpha$ and $\beta$, and compare this with the
moduli space metric of the BCFT derived in
section
\ref{s3}. 
For this we treat the parameters $\lambda$, $\alpha$ and $\beta$ as
collective coordinates, and consider a time dependent configuration of the
form:
\be \label{ess7}
T(x,t) = \sqrt 2 \sinh^{-1} \left({\lambda(t)\over \sqrt 2} 
\cos\left({x\over 
\sqrt
2} +
\alpha(t)\right)\right), \quad A_x(x,t) = {1\over \sqrt 2} \beta(t)\, .
\ee
Substituting this into the action \refb{ess1a} we get the action of the 
collective coordinates. A short calculation 
gives:
\ben \label{et1}
S &=& -2\pi \sqrt 2 \, \TT_1\, \Bigg[1- \left( {1\over 
\sqrt{1+{1\over 2} \lambda^2}} -  
{1\over 1+{1\over 2}\lambda^2}\right)\, {\dot\lambda^2\over \lambda^2} -
\left( 1 - 
{1\over
\sqrt{1+{1\over 2}\lambda^2}}\right) \dot\alpha^2 \nonumber \\
&& \qquad - {1\over 
4}\, {1\over
\sqrt{1+{1\over 2}\lambda^2}}\, 
\dot\beta^2
\Bigg] + \ldots\, ,
\een
where $\ldots$ denotes terms involving four or more time derivatives.
This gives the moduli space metric to be
\be \label{et2}
ds^2 = 2\left[\left( {1\over
\sqrt{1+{1\over 2}\lambda^2}} -
{1\over 1+{1\over 2}\lambda^2}\right)\, {d\lambda^2\over \lambda^2} + 
\left( 1 -
{1\over
\sqrt{1+{1\over 2}\lambda^2}}\right) d\alpha^2 + {1\over 4}\, {1\over
\sqrt{1+{1\over 2}\lambda^2}}\, 
d\beta^2\right]\, .
\ee

In order to compare this with the metric \refb{e2.6} found from direct 
BCFT analysis, we
need to find the relation between the
parameters $(\lambda,\alpha,\beta)$ appearing in \refb{ess2} and the
parameters $(\theta,\phi,\chi)$ labelling the BCFT. For this we note that 
for small $\lambda$, the solution given in
\refb{ess2} reduces to the configuration \refb{e2.11}. Thus for small
$\lambda$ we can use the results of eqs.\refb{e2.15} to get
\be \label{ess4}
\alpha=\phi, \quad \beta = 2\chi, \quad \lambda \simeq 2 
\theta\, .
\ee
We expect that for the angular 
variables
$\chi$, $\phi$, $\alpha$ and $\beta$ this identification will continue to 
hold even for 
finite $\lambda$ since in the effective
field theory (conformal field theory) $\alpha$ ($\phi$) has 
the
interpretation of translation along $S^1$ and $\beta/2$ ($\chi$) 
has the
interpretation of switching on constant gauge field. The relation between
$\lambda$ and $\theta$ however is likely to be modified for finite 
$\lambda$. To determine this
relation we use the same principle as was used in section \ref{s3}, --
namely compare the $T_{xx}$ associated with these
solutions. For the
BCFT we have
\be \label{ess5}
T_{xx} = - \TT_1 \cos^2(\pi\tl) = - \TT_1 \cos^2\theta\, ,
\ee
whereas for the solution \refb{ess2} and the action \refb{ess1a}
\be \label{ess6}
T_{xx} = - {V(T) \over \sqrt{1 +(T')^2}} = - {\TT_1 \over
\sqrt{1+{1\over 2}\lambda^2}}\, .
\ee
Comparing \refb{ess5} and \refb{ess6} we get:
\be \label{ess6a}
{1\over 1+{1\over 2}\lambda^2} = \cos^4\theta\, .
\ee
Thus the complete relation between the parameters $(\lambda,\alpha,\beta)$ 
and $(\theta,\phi,\chi)$ is given by:
\be \label{ess6b}
\lambda = \sqrt 2\, \sec^2\theta\sin\theta\sqrt{1+\cos^2\theta}, \quad
\alpha=\phi, 
\quad \beta=2\chi\, .
\ee
Note that as $\lambda$ varies between 0 and $\infty$, $\theta$ varies 
between 0 and $\pi/2$. On the other hand since $\alpha$ and $\beta$ 
are angular variables with periodicity $2\pi$, $\phi$ and $\chi$ are angular 
variables with periodicities $2\pi$ and $\pi$ respectively. Thus the 
ranges of 
$(\theta,\phi,\chi)$ are given by:
\be \label{eex2}
0\le\theta\le {\pi\over 2}, \quad 0\le\phi\le 2\pi, \quad 0\le \chi\le
\pi\, .
\ee
This is identical to \refb{eextra} obtained by analysis of the moduli 
space of the BCFT.\footnote{The fact that the ranges of $\phi$ and $\chi$
agree between effective field theory and boundary conformal field theory 
is not a surprise since the relative normalization between $\phi$ and 
$\alpha$ ($\chi$ and $\beta$) was fixed by matching their periodicities. 
Matching of the range of $\theta$ however was not guaranteed.} 

Using \refb{ess6b} we can express the metric \refb{et2} in terms of the 
variables $(\theta,\chi,\phi)$. The result is:
\be \label{ess9}
ds^2 = 2\left[ {4 d\theta^2 \over (1 +
\cos^2\theta)^2} + \sin^2\theta d\phi^2 + \cos^2\theta
d\chi^2\right]\, ,
\ee
This is to be compared with the exact answer \refb{e2.6}. We note first of 
all that the two metrics agree near $\theta=0$. This is not surprising 
since the effective action \refb{ess1a} agrees with the
string field theory action 
\refb{e2.7} for small $T$ and $A_x$, and \refb{e2.7} does describe 
correctly the dynamics of small fluctuations on a non-BPS D-brane. 
However the similarity between \refb{e2.6} and 
\refb{ess9} goes beyond the small $\theta$ region. The coefficients of 
$d\phi^2$ and $d\chi^2$ terms are reproduced exactly. Furthermore the 
ranges of $(\theta,\phi,\chi)$ labelling the solutions of the effective 
field theory and BCFT's are identical. 
The only 
difference between the two metrics is in the coefficient of the 
$d\theta^2$ term. For any given $\theta$ the coefficient of the 
$d\theta^2$ term is larger in the metric \refb{ess9} derived from the 
effective field 
theory than in the exact metric \refb{e2.6}. Thus the moduli space of the 
solutions of 
the effective field theory can be regarded as an elongated version of the 
exact moduli space $S^3/Z_2$ along the $\theta$ coordinate.

To get more insight into the geometric stucture of the moduli space 
described by the metric \refb{ess9} we go to the double cover by taking 
$\chi$ to be a periodic variable with periodicity 
$2\pi$. We can now consider a 
family of metrics interpolating between the metric \refb{ess9} and the 
metric on $S^3$:
\be \label{ess10}
ds^2 = 2\left[ { 4 d\theta^2 \over (2 -
u\sin^2\theta)^2} + \sin^2\theta d\phi^2 + \cos^2\theta
d\chi^2\right]\, .
\ee
As $u$ varies from 0 to 1, we go from the metric on $S^3$ to the metric 
\refb{ess9}. For a given $\theta$ the coefficient of $d\theta^2$ increases 
monotonically as $u$ increases. Thus as $u$ varies from 0 to 
1, \refb{ess10} describes a family of metrics in which $S^3$ is elongated 
continuously along the $\theta$ direction. 

Near $\theta=\pi/2$ the metric \refb{ess10} looks like
\be \label{ess11}
ds^2 \simeq 2\left[d\phi^2 + {4\over (2-u)^2} d\psi^2 + \psi^2 
d\chi^2\right] \, ,
\ee
where $\psi = {\pi\over 2} - \theta$. This metric has a conical defect 
with deficit angle $u\pi$ at $\theta=\pi$. Thus in the process of 
elongating $S^3$ along the $\theta$ direction, the metric \refb{ess10} 
also develops a conical singularity at $\theta=\pi/2$ where the excess 
curvature is stored. For the metric \refb{ess9} the deficit angle at 
$\theta=\pi/2$ reaches the value $\pi$, {\it i.e.} we have a $Z_2$ 
orbifold 
singularity at this point. The 
actual moduli space is obtained by taking a further quotient of this 
manifold by $Z_2$, which generates a $Z_2$ orbifold singularity on the 
original $S^3$ and converts the $Z_2$ orbifold singularity of the metric 
\refb{ess9} into a $Z_4$ orbifold singularity.

At $\theta=\pi/2$ ($\lambda=\infty$) the solution \refb{ess2} represents 
infinitely sharp kink antikink pair located at $x=\sqrt 2({\pi\over 2} 
-\alpha)$ and $x=\sqrt 2({3\pi\over 2}
-\alpha)$. Thus at this point a new branch of moduli space opens up, 
consisting of configurations where the kink and the antikink are placed at 
arbitrary points on the circle. The moduli space of these solutions, 
labelled by the locations of the kink and the antikink on $S^1$, 
clearly has the structure of a square torus $S^1\times S^1$, and is joined 
to the other branch along the diagonal where the kink and the anti-kink 
are placed at diametrically opposite points. This is precisely what 
happens for the exact moduli space obtained from the BCFT analysis. In 
fact, in this case one can show that the full effective action living on 
the world-volume of the kink-antikink pair is given by the 
Dirac-Born-Infeld action on a D0-$\bd0$ brane pair\cite{effective3}. Thus 
the correspondence between the BCFT description and the world-volume 
description of the action goes far beyond the moduli space approximation. 

In order to have a new branch opening up at the $\theta={\pi\over 2}$ 
point, there must be additional massless modes corresponding to changing 
the relative separation between the kink and the antikink. Thus we should 
expect that when we analyze the spectrum of fluctuations around the 
background \refb{ess2}, a new massless mode, besides the three collective 
coordinates, must appear in the $\lambda\to\infty$ limit. In section 
\ref{s5} we verify this explicitly.

To summarize, we see that qualitatively the structure of the moduli space
of classical solutions in the effective field theory described by the
action \refb{ez1} is in agreement with the exact result obtained in
section \ref{s3} using the bondary conformal field theory analysis. This
shows that the effective field theory \refb{ez1} is a good candidate for
providing a qualitative description of tachyon dynamics on an unstable
D-brane.

\sectiono{Origin of the Additional Massless Mode at the Dirichlet Point} 
\label{s5}

In the analysis of sections \ref{s3} and \ref{s4} we have seen that at 
the point $\theta=\pi/2$, additional massless modes appear, and a new 
branch of the moduli space opens up. The origin of these additional 
massless modes in the BCFT is well 
understood\cite{9808141,9812031}, -- the usual zero momentum tachyonic 
mode and various other massive 
modes at the $\theta=0$ point become massless at $\theta=\pi/2$ and give 
rise 
to the new branch of the moduli space. In this section we shall explore 
the origin of the additional massless mode in the effective field theory 
description of section \ref{s4}. In particular we shall show that in the 
effective field 
theory also the zero momentum tachyonic mode at the $\theta=0$ 
point flows 
into a massless mode at the $\theta=\pi/2$ point, and gives rise to the 
additional direction in the moduli space that allows us to change the 
relative separation between the kink antikink pair.

We shall begin by analyzing the small oscillation modes 
around 
the classical solution \refb{ess2}. Due to the shift symmetry $A_x\to 
A_x+c$, and the translational invariance along $x$ of the effective field 
theory action \refb{ess1a}, we can work at the point $\alpha=\beta=0$ 
without any loss of generality. Thus the solution is:
\be \label{e5.1}
T = \sqrt 2 \sinh^{-1} \left({\lambda\over \sqrt 2} \cos\left({x\over
\sqrt 2}\right)\right), \quad A_x = 0 \, .
\ee 
In carrying out the analysis of fluctuation around this solution, we shall 
ignore the fluctuation of the gauge 
field $A_x$ since, as we shall see, the additional massless mode at 
$\theta=\pi/2$ appears from the tachyon field $T$. This is a consistent 
truncation due to $A_x\to -A_x$ 
symmetry of the action. Thus we focus 
our attention on the following type 
of fluctuation around the classical solution \refb{e5.1}
\be \label{e5.2}
T(x,t) = \sqrt 2 \sinh^{-1} \left({\lambda\over \sqrt 2} \cos\left({x\over
\sqrt 2}\right)\right) + \chi(x, t), \quad A_x = 0 \, .
\ee
Substituting this into the effective action \refb{ess1a}, and ignoring the 
constant term, we get the 
following action for the mode $\chi$ to quadratic order in $\chi$:
\ben \label{e5.3}
\SSS &=& {\TT_1\over \sqrt{4 +2\lambda^2}} \int dt \int_0^{2\pi\sqrt 2} 
dx 
\left[\dot\chi^2 - {2+\lambda^2 \cos^2 (x/\sqrt 2)\over 2 + 
\lambda^2} 
(\chi')^2
+ {1\over 2+\lambda^2 \cos^2 (x/\sqrt 2)} \chi^2\right] \nonumber \\
&& + 
\OO(\chi^3)\, .
\een
Thus the linearized equation of motion for $\chi$ is:
\be \label{e5.4}
-{d\over dx} \left[  {2+\lambda^2 \cos^2 (x/\sqrt 2)\over 2 + 
\lambda^2} 
{d\chi\over 
dx} \right] - {1\over  2+\lambda^2 \cos^2 (x/\sqrt 2)} \chi = -\ddot\chi\, 
.
\ee
A mode of $\chi$ of mass $m$ will correspond to a time dependent solution 
of the form $\chi_m(x) e^{imt}$. Substituting this into \refb{e5.4} we 
get:
\be \label{e5.5} 
-{d\over dx} \left[  {2+\lambda^2 \cos^2 
(x/\sqrt 2)\over 2 +
\lambda^2}
{d\chi_m\over
dx} \right] - {1\over  2+\lambda^2 \cos^2 (x/\sqrt 2)} \chi_m = 
m^2\chi_m\, .
\ee
Using the relation between $\lambda$ and the parameter $\theta$ given in 
\refb{ess6a} we can rewrite this equation as:
\ben \label{e5.5a}
&& -{d\over dx} \left[  \left(\cos^4\theta + \sin^2\theta (1 + 
\cos^2\theta) \cos^2 {x\over \sqrt 2} \right)
{d\chi_m\over
dx} \right] \nonumber \\
&& \qquad - {\cos^4\theta\over  2\left(\cos^4\theta +\sin^2\theta 
(1 
+\cos^2\theta) \cos^2 (x/\sqrt 2)\right)} \chi_m =
m^2 \chi_m\, .
\een 
This is an eigenvalue equation for the operator 
\be \label{e5.6}
-{d\over dx} \left[  \left(\cos^4\theta + \sin^2\theta (1 + 
\cos^2\theta) \cos^2 {x\over \sqrt 2} \right)
{d\over
dx} \right] 
- {\cos^4\theta\over  2(\cos^4\theta +\sin^2\theta 
(1 
+\cos^2\theta) \cos^2 (x/\sqrt 2))} 
\ee
with the eigenvalues giving the mass$^2$ spectrum of the $(0+1)$ 
dimensional theory living on this classical solution. For any value of 
$\theta$ this operator contains a zero 
eigenvalue, with the corresponding eigenfunction $\chi_0$ denoting 
deformation induced by the parameter $\alpha$ in eq.\refb{ess2}:
\be \label{e5.7}
\chi_0(x) \propto {d\over dx} \left[ \sinh^{-1} \left({\lambda\over \sqrt 
2} \cos\left({x\over
\sqrt 2}\right)\right) \right] \propto {\sin(x/\sqrt 2) \over \sqrt{1 + 
\tan^2\theta (1 + \sec^2\theta) \cos^2 (x/\sqrt 2)}}\, .
\ee
It can be checked explicitly that $\chi_0$ defined in \refb{e5.7} 
satisfies \refb{e5.5a} with $m=0$.

\begin{figure}[!ht]
\leavevmode
\begin{center}
\epsfysize=5cm
\epsfbox{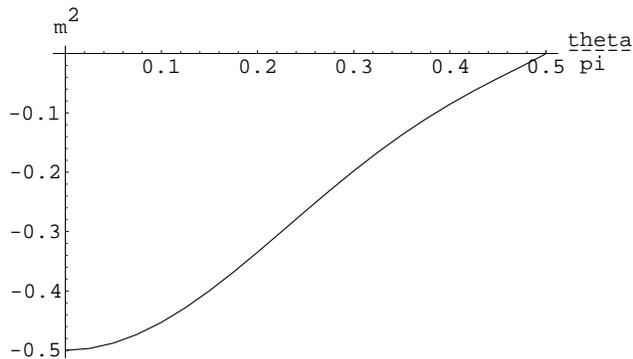}
\end{center}
\caption{Numerical results for the mass$^2$ of the lowest mode of the 
tachyon as a function of the parameter $\theta/\pi$ labelling the 
classical 
solution in the effective field theory. Note that the mass$^2$ goes from 
$-{1\over 2}$ to 0 as the parameter $\theta$  varies from 0 to $\pi/2$.} 
\label{f1}
\end{figure}

{}From the form of $\chi_0$ given in \refb{e5.7} we see that it has two 
nodes in the range $0\le x< 2\pi\sqrt 2$, -- at $x=0$ and at $x=\pi\sqrt 
2$. 
Arguments given in \cite{0304197} then show that there must be another 
eigenfunction of the operator \refb{e5.6} of lower $m^2$ eigenvalue, ${\it 
i.e.}$ with $m^2 \le 0$. Such an eigenfunction
clearly exists at $\theta = 0$, -- it is simply the constant mode 
and has $m^2 = -{1\over 2}$. For $\theta\ne 0$, the eigenfunction and the 
eigenvalue can be computed numerically. In Fig.\ref{f1} we show the 
numerical results for the lowest $m^2$ eigenvalue as a function of 
$\theta/\pi$. As is clear from this figure, as $\theta$ varies from 0 to 
$\pi/2$, the lowest $m^2$ value varies smoothly from $-1/2$ to 0. This is 
consistent with the general argument of \cite{0304197} that the tachyonic 
mode must disappear in the $\lambda\to\infty$ ({\it i.e.} $\theta\to 
\pi/2$) limit.

\begin{figure}[!ht]
\leavevmode
\begin{center}
\epsfysize=5cm
\epsfbox{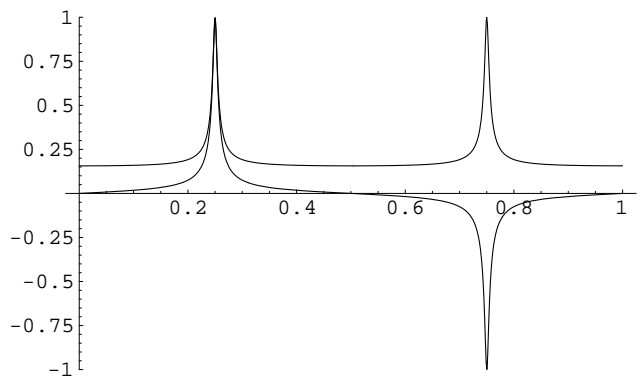}
\end{center}
\caption{Plot of $\chi$ vs. $x/(2\pi\sqrt 2)$ for the two lowest modes for 
$\theta=.45 \pi $. The lowest mode has positive peaks at $x=\pi/2$ and 
at $x=3\pi / 2$, whereas the first excited mode (translational zero mode) 
has a positive peak at $x=\pi/2$ and a negative peak at $x=3\pi/2$. } 
\label{f2} \end{figure}

In order to find the physical interpretation of this mode in the 
$\theta\to \pi/2$ limit, we can examine the form of the 
eigenfunction. In fig.~\ref{f2} we have plotted the suitably normalized 
eigenfunction as a 
function of $(x/2\pi\sqrt 2)$ for $\theta = .45 \pi $. For comparison, we 
have also plotted the translational mode 
given in \refb{e5.7} on the same graph. Comparing the two 
graphs we see that both are peaked around the points $x=\pi / \sqrt 2$, 
and $3\pi/\sqrt 2$, {\it i.e.} the points where the kink and the 
antikink are located in the $\theta\to\pi/2$ limit. Furthermore the 
forms of the lowest mode and the translational mode agree closely near 
$x = \pi / \sqrt 2$, but differ by a sign near $x=3\pi/\sqrt 2$. 
Numerical results also show that for both eigenmodes the peaks become 
sharper as $\theta$ approaches $\pi/2$.
Since 
at $\theta=\pi/2$ the translational mode corresponds to translating the 
kink
and the antikink in the same direction, keeping their relative 
separation intact, this suggests that the lowest mode corresponds to 
translating the kink and the antikink in the opposite direction. This 
is precisely the mode that takes us to the other branch of the moduli 
space.

{\bf Acknowledgement}: I wish to thank R.~Gopakumar for useful 
discussions.


\begin{thebibliography}{99}

\bibitem{others}
C.G.~Callan, I.R.~Klebanov, A.W.~Ludwig and J.M.~Maldacena,
Nucl. Phys. {\bf B422}, 417 (1994)
hep-th/9402113; \\
J.~Polchinski and L.~Thorlacius,
Phys. Rev. {\bf D50}, 622 (1994)
hep-th/9404008; \\
A.~Sen,
Int.\ J.\ Mod.\ Phys.\ A {\bf 14}, 4061 (1999)
[arXiv:hep-th/9902105].


\bibitem{9811237}
A.~Recknagel and V.~Schomerus,
Nucl.\ Phys.\ B {\bf 545}, 233 (1999)
[arXiv:hep-th/9811237]; \\
M.~R.~Gaberdiel, A.~Recknagel and G.~M.~T.~Watts,
Nucl.\ Phys.\ B {\bf 626}, 344 (2002)
[arXiv:hep-th/0108102].

\bibitem{9808141}
A.~Sen,
JHEP {\bf 9809}, 023 (1998)
[arXiv:hep-th/9808141].

\bibitem{9812031}
A.~Sen,
JHEP {\bf 9812}, 021 (1998)
[arXiv:hep-th/9812031].

\bibitem{0108238}
M.~R.~Gaberdiel and A.~Recknagel,
JHEP {\bf 0111}, 016 (2001)
[arXiv:hep-th/0108238].


\bibitem{0203211}
A.~Sen,
JHEP {\bf 0204}, 048 (2002)
[arXiv:hep-th/0203211].

\bibitem{0203265}
A.~Sen,
JHEP {\bf 0207}, 065 (2002)
[arXiv:hep-th/0203265].

\bibitem{0207105}
A.~Sen,
JHEP {\bf 0210}, 003 (2002)
[arXiv:hep-th/0207105].

\bibitem{larsen}
F.~Larsen, A.~Naqvi and S.~Terashima,
JHEP {\bf 0302}, 039 (2003)
[arXiv:hep-th/0212248]; \\
N.~R.~Constable and F.~Larsen,
arXiv:hep-th/0305177.

\bibitem{rolling}
M.~Gutperle and A.~Strominger,
JHEP {\bf 0204}, 018 (2002)
[arXiv:hep-th/0202210]; \\
A.~Strominger,
arXiv:hep-th/0209090; \\
M.~Gutperle and A.~Strominger,
arXiv:hep-th/0301038; \\
A.~Maloney, A.~Strominger and X.~Yin,
arXiv:hep-th/0302146; \\
P.~Mukhopadhyay and A.~Sen,
JHEP {\bf 0211}, 047 (2002)
[arXiv:hep-th/0208142]; \\
T.~Okuda and S.~Sugimoto,
Nucl.\ Phys.\ B {\bf 647}, 101 (2002)
[arXiv:hep-th/0208196]; \\
N.~Moeller and B.~Zwiebach,
JHEP {\bf 0210}, 034 (2002)
[arXiv:hep-th/0207107]; \\
M.~Fujita and H.~Hata,
arXiv:hep-th/0304163; \\
I.~Y.~Aref'eva, L.~V.~Joukovskaya and A.~S.~Koshelev,
arXiv:hep-th/0301137; \\
S.~J.~Rey and S.~Sugimoto,
Phys.\ Rev.\ D {\bf 67}, 086008 (2003)
[arXiv:hep-th/0301049]; \\
S.~J.~Rey and S.~Sugimoto,
arXiv:hep-th/0303133; \\
J.~L.~Karczmarek, H.~Liu, J.~Maldacena and A.~Strominger,
arXiv:hep-th/0306132; \\
V.~Schomerus,
arXiv:hep-th/0306026; \\
S.~Fredenhagen and V.~Schomerus,
arXiv:hep-th/0308205.


\bibitem{effective}
A.~Sen,
JHEP {\bf 9910}, 008 (1999)
[arXiv:hep-th/9909062]; \\
M.~R.~Garousi,
Nucl.\ Phys.\ B {\bf 584}, 284 (2000)
[arXiv:hep-th/0003122], 
JHEP {\bf 0305}, 058 (2003)
[arXiv:hep-th/0304145]; \\
E.~A.~Bergshoeff, M.~de Roo, T.~C.~de Wit, E.~Eyras and S.~Panda,
JHEP {\bf 0005}, 009 (2000)
[arXiv:hep-th/0003221]; \\
A.~Sen,
Mod.\ Phys.\ Lett.\ A {\bf 17}, 1797 (2002)
[arXiv:hep-th/0204143].

\bibitem{0303139}
N.~Lambert, H.~Liu and J.~Maldacena,
arXiv:hep-th/0303139.

\bibitem{effective2}
K.~Kamimura and J.~Simon,
Nucl.\ Phys.\ B {\bf 585}, 219 (2000)
[arXiv:hep-th/0003211]; \\
J.~Kluson,
Phys.\ Rev.\ D {\bf 62}, 126003 (2000)
[arXiv:hep-th/0004106]; \\
G.~W.~Gibbons, K.~Hori and P.~Yi,
Nucl.\ Phys.\ B {\bf 596}, 136 (2001)
[arXiv:hep-th/0009061]; \\
A.~Sen,
arXiv:hep-th/0209122; \\
J.~A.~Minahan and B.~Zwiebach,
JHEP {\bf 0102}, 034 (2001)
[arXiv:hep-th/0011226]; \\
M.~Alishahiha, H.~Ita and Y.~Oz,
Phys.\ Lett.\ B {\bf 503} (2001) 181
[arXiv:hep-th/0012222]; \\
C.~j.~Kim, H.~B.~Kim, Y.~b.~Kim and O.~K.~Kwon,
arXiv:hep-th/0301076; \\
F.~Leblond and A.~W.~Peet,
arXiv:hep-th/0303035; \\
D.~Kutasov and V.~Niarchos,
arXiv:hep-th/0304045; \\
M.~Smedback,
arXiv:hep-th/0310138; \\
G.~Gibbons, K.~Hashimoto and P.~Yi,
JHEP {\bf 0209}, 061 (2002)
[arXiv:hep-th/0209034]; \\
O.~K.~Kwon and P.~Yi,
arXiv:hep-th/0305229; \\
A.~Buchel, P.~Langfelder and J.~Walcher,
Annals Phys.\  {\bf 302}, 78 (2002)
[arXiv:hep-th/0207235]; \\
G.~N.~Felder, L.~Kofman and A.~Starobinsky,
JHEP {\bf 0209}, 026 (2002)
[arXiv:hep-th/0208019]; \\
K.~Hashimoto and S.~Hirano,
Phys.\ Rev.\ D {\bf 65}, 026006 (2002)
[arXiv:hep-th/0102174]; \\
N.~D.~Lambert and I.~Sachs,
Phys.\ Rev.\ D {\bf 67}, 026005 (2003)
[arXiv:hep-th/0208217]; \\
A.~Sen,
arXiv:hep-th/0305011; \\
A.~Fotopoulos and A.A.~Tseytlin, arXiv:hep-th/0310253; \\
E.~J.~Copeland, P.~M.~Saffin and D.~A.~Steer,
Phys.\ Rev.\ D {\bf 68}, 065013 (2003)
[arXiv:hep-th/0306294].

\bibitem{effective3}
N.~D.~Lambert and I.~Sachs,
JHEP {\bf 0106}, 060 (2001)
[arXiv:hep-th/0104218]; \\
G.~Arutyunov, S.~Frolov, S.~Theisen and A.~A.~Tseytlin,
JHEP {\bf 0102}, 002 (2001)
[arXiv:hep-th/0012080]; \\
K.~Hashimoto and S.~Nagaoka,
Phys.\ Rev.\ D {\bf 66}, 026001 (2002)
[arXiv:hep-th/0202079]; \\
A.~Sen,
Phys.\ Rev.\ D {\bf 68}, 066008 (2003)
[arXiv:hep-th/0303057].


\bibitem{FMS}
D.~Friedan, E.~J.~Martinec and S.~H.~Shenker,
Nucl.\ Phys.\ B {\bf 271}, 93 (1986).


\bibitem{inverse}
E.~Witten,
Nucl.\ Phys.\ B {\bf 276} (1986) 291.

\bibitem{0304197}
P.~Brax, J.~Mourad and D.~A.~Steer,
arXiv:hep-th/0304197.

\end{thebibliography}
\end{document}